\documentclass[aps, superscriptaddress, ctexart, nofootinbib, twocolumn]{revtex4-1}

\pdfoutput=1

\usepackage{amssymb, amsmath, bm, dcolumn, graphicx, latexsym, slashed, simplewick}
\usepackage[utf8]{inputenc}

\usepackage{color}

\def\be{\begin{equation}}
\def\ee{\end{equation}}
\def\bea{\begin{eqnarray}}
\def\eea{\end{eqnarray}}

\input epsf.tex
\epsfclipon

\usepackage{epsfig}
\usepackage{epstopdf}
\usepackage{epsf}
\input{epsf.sty}

\begin{document}

\title{The Cosmological Heavy Ion Collider: Fast Thermalization after Cosmic Inflation}

\author{Evan McDonough}
\affiliation{Brown Theoretical Physics Center and Department of Physics, Brown University, 182 Hope Street, Providence, RI. 02903}
\affiliation{Center for Theoretical Physics,\\
Massachusetts Institute of Technology, Cambridge, MA 02139, USA}

\begin{abstract}

Heavy-ion colliders have revealed the process of ``fast thermalization''. This experimental break-through has led to new theoretical tools to study the thermalization process at both weak and strong coupling.  We apply this to the reheating epoch of inflationary cosmology, and the formation of a cosmological quark gluon plasma (QGP). We compute the thermalization time of the QGP at reheating, and find it is determined by the energy scale of inflation and the shear viscosity to entropy ratio $\eta/s$; or equivalently, the tensor-to-scalar ratio and the strong coupling constant at the epoch of thermalization. Thermalization is achieved near-instantaneously in low-scale inflation and in strongly coupled systems, and takes of order or less than a single e-fold of expansion for weakly-coupled systems or after high-scale inflation. We demonstrate that the predictions of inflation are robust to the physics of thermalization, and find a stochastic background of gravitational waves at frequencies accessible by interferometers, albeit with a small amplitude.
 \end{abstract}

\maketitle

\section{Introduction}

Thermalization in quantum field theory has a long history (see e.g.~\cite{Horsley:1985dz}), and, in particular, thermalization in non-Abelian gauge theory has been a driving research question in quantum chromodynamics (QCD) for decades; for recent reviews, see \cite{Schlichting:2019abc,Busza:2018rrf}. There are two places in the history of the universe when a non-Abelian gauge theory at high density has thermalized: at the present day at the Relativistic Heavy Ion Collider (RHIC) \cite{Shuryak:2004cy,Heinz:2004pj}\footnote{In this work the distinction between the RHIC observation of thermalization vs.  hydrodynamization will not be relevant; for a discussion of this topic see, e.g., \cite{Nagle:2018nvi, Romatschke:2016hle}.}, and in the early universe, at the beginning of the radiation dominated era of standard cosmology.

A primary question that has arisen in the QCD community has been the mechanism behind the rapid approach to thermal equilibrium, so-called \emph{fast thermalization}. This has led to radically different approaches, both at weak coupling \cite{Arnold:2002zm, Arnold:2003zc, Arnold:2004ti,Arnold:2000dr,Jeon:1995zm}, where a particle description can be applied, and at strong coupling \cite{Kovtun:2004de, Bhattacharyya:2008jc, Chesler:2007sv, Kovtun:2005ev}, where thermalization corresponds to black hole formation in an anti-de Sitter spacetime \cite{Balasubramanian:2011ur}. 

Parallel to this, a central question in early universe cosmology, which has risen in prominence with the precision of cosmic microwave background experiments \cite{Akrami:2018vks}, is the duration of the period of \emph{reheating} that necessarily follows cosmic inflation \cite{Dolgov:1982th,*Abbott:1982hn,*Albrecht:1982mp,Traschen:1990sw,Shtanov:1994ce,Kofman:1994rk,Kofman:1997yn}; for reviews see e.g.~\cite{Allahverdi:2010xz,Amin:2014eta,Lozanov:2019jxc,Lozanov:2019jff}. The uncertainty in the duration of reheating leads to an uncertainty in two of the hallmark predictions of inflationary cosmology, namely the scalar spectral index $n_s$ and the tensor-to-scalar ratio $r$ \cite{Martin:2010kz,Munoz:2014eqa,Gong:2015qha,Cook:2015vqa} of the primordial perturbations.  Narrowing and quantifying these uncertainties has been a primary aim of the reheating literature \cite{Dai:2014jja,Martin:2016oyk,Drewes:2015coa,Lozanov:2016hid,Ji:2019gfy,Kabir:2016kdh}; and has spurred on detailed numerical calculations \cite{Adshead:2015pva,Nguyen:2019kbm,Ema:2016dny,Lozanov:2016pac,Iarygina:2018kee,Sfakianakis:2018lzf,Subramanian:2015lua,DeCross:2016fdz,vandeVis:2020qcp}.

This issue aside, the self-consistency of the inflationary universe scenario requires the universe reach a state of near thermal equilibrium at a temperature of or above $\sim$MeV \cite{Kawasaki:2000en,Hannestad:2004px,Hasegawa:2019jsa}, else a signature success of standard cosmology, namely big bang nucleosynthesis, would be forgone. This imposes a constraint  that the universe thermalize before the energy density of the universe redshifts below MeV$^4$, providing a baseline test of self-consistency, which one may intuit to be satisfied, but ultimately must be derived or numerically demonstrated.

It is conventional in the cosmology community to define the  transition from reheating to standard cosmology as the moment when the universe is dominated by relativistic particles, or alternatively, when the equation of state of the universe is sufficiently close to $w=1/3$, see e.g. \cite{Lozanov:2016hid}. However, by ignoring the thermalization process, this abandons the rich underlying physics, and potentially rich phenomenology, that can be accessed by applying the machinery developed for thermalization at heavy ion colliders.

In this work we will study the thermalization process during the reheating epoch after cosmic inflation, which has been neglected in all but a few previous works \cite{Brandenberger:2019njw,Micha:2002ey,Micha:2004bv,Mukaida:2015ria,Garcia:2018wtq,Harigaya:2019tzu,Harigaya:2013vwa,Kawai:2015lja}\footnote{There are also but few works that apply quark gluon plasma physics to cosmology generally \cite{Sanches:2014gfa,Ghiglieri:2015nfa,Rangamani:2015qga}.}. We apply the formalism and results of the thermalization literature, namely the relation between thermalization time, temperature, and shear-viscosity-to-entropy ratio $\eta/s$, and the theoretical predictions for $\eta/s$ at strong and weak coupling. From this we will demonstrate that a post-inflationary standard model plasma undergoes \emph{fast thermalization}. We will compute the imprint on cosmological observables, namely on $n_s$, $r$, and the production of a stochastic background of gravitational waves. While the imprint is small, this sets the stage for further applications of modern QCD techniques to the very early universe.

\section{Thermalization at the Cosmological Heavy Ion Collider}

\label{sec:thermFRW}

There are many theoretical approaches to the quark gluon plasma, with different regimes of applicability, depending on the coupling and occupation number of the plasma.  At weak coupling and perturbative occupation numbers $n \ll 1/\alpha_s$, one can derive an effective kinetic theory governed by a Boltzmann equation \cite{Baier:2000sb,Mueller:2002gd,Arnold:2000dr}.  At weak coupling but non-perturbative occupation numbers $n_k \sim 1/\alpha_s$, one can take a different approach and model the system as weakly coupled fluctuations of a condensate \cite{Iancu:2003xm}, a so-called {\it Color-Glass Condensate}. In this same regime, lattice gauge theory studies suggest that there may occur a process of {\it Turbulent Thermalization} \cite{Berges:2013fga}. For couplings that are large but still smaller than the non-perturbative value $1/\alpha_s$, the system is effectively classical and perturbative calculations can be made in classical Yang-Mills theory \cite{Kurkela:2012hp}. 

A general result of these approaches is that the thermalization time and temperature of the quark gluon plasma  are both related to the shear viscosity to entropy ratio. The three quantities are intertwined as \cite{Keegan:2015avk,vanderSchee:2017edo},
\be
\label{eq:tre}
t_{th} \simeq \left(\frac{\eta}{s} \right)\frac{1}{T_{th}},
\ee
where  $\eta/s$ is the shear viscosity to entropy ratio. This is in turn determined by the interaction strength of the quantum field theory.

At strong coupling, the only analytical tool available is holography, namely the Anti-de Sitter/Conformal Field Theory correspondence (AdS/CFT) \cite{Maldacena:1997re}. In this context the thermalization of a non-Abelian plasma can be described as black hole formation in an anti-de Sitter spacetime \cite{Balasubramanian:2011ur}. A remarkable result from this field is a lower bound on the shear-viscosity-to-entropy ratio \cite{Policastro:2001yc,Kovtun:2005ev,Cremonini:2011iq},
\be
\label{etasbound}
\frac{\eta}{s} \gtrsim \frac{1}{4 \pi}. 
\ee
This is computed in the gravity dual as the zero-frequency limit of the graviton absorption cross-section of black three-branes.  This lower bound can be violated by higher-derivative terms in the gravitational action \cite{Brigante:2007nu}, or by the inclusion of bulk viscosity. For the purposes of this work, we consider the above to be a lower bound.

In what follows, we will extend \eqref{eq:tre} to an FRW universe. We will work in the simple approximation that the inflaton has decayed completely into particles, e.g. after rapid reheating. This will allow us to calculate the number of e-folds of the thermalization process after inflation.  Importantly, we note that \eqref{eq:tre} is independent of the expansion of space: solving self-consistently for the redshifting of energy and the thermalization time alters the scaling with $\eta/s$ of both $t_{th}$ and $T_{th}$, leaving \eqref{eq:tre} unchanged. 

The expansion of space is related to the thermalization temperature via the Friedmann equation. That is, the thermalization temperature is related to the Hubble parameter at the end of thermalization as,
\be
H_{th}^2 = \frac{\pi^2}{30} g_*\frac{T_{th}^4}{3 m_{pl}^2} 
\label{HreTre}
\ee
where $g_*$ is the number of relativistic degrees of freedom. Meanwhile,  the time evolution of the Hubble constant, for an FRW universe with equation of state $w>-\frac{1}{3}$, is given by
\be
H(t) = \frac{2}{3(1+w)t} = \frac{H_i}{1 + \hat{w} H_i \Delta t},
\ee
where the latter equality applies for a constant $w$,  $\Delta t \equiv t-t_i$, and $\hat{w}$ is defined as
\be
\label{hatw}
\hat{w} \equiv \frac{3}{2}(1+w) \;\; , \;\; 1\leq \hat{w} \leq 2.
\ee 
Thus the Hubble constant at the end of thermalization can also be written as
\be
H_{th} = \frac{H_i}{1 + \hat{w} H_i t_{th}}.
\label{Hretre}
\ee
Combining equations \eqref{HreTre} and \eqref{Hretre} with \eqref{eq:tre} uniquely fixes $t_{th}$ and $T_{th}$ in terms of $\eta/s$, $w$, and the initial value of the Hubble constant $H_i$. 

The Hubble constant at the culmination of thermalization is,
\begin{eqnarray}
\label{Hth}
\frac{H_{th}}{H_i} = 1\,+\,&& \frac{\sqrt{g_*} \pi}{6\sqrt{10}} \left(\frac{\hat{w}\eta}{s} \right)^2 \left(\frac{H_i}{m_{pl}} \right) \\
&& \cdot \Bigg[ 1 -\sqrt{1 + \frac{12 \sqrt{10}m_{pl}}{(\hat{w} \eta/s)^2 \sqrt{g_*} \pi H_i}} \;\Bigg]. \nonumber
\end{eqnarray}
 From this we see that the equation of state $w$ acts as a simple rescaling of $\eta/s$. One can also see that $H_{th} \rightarrow H_i $ at strong coupling, i.e.~in the limit $\eta/s \rightarrow 0$, while at weak coupling, $\eta/s\rightarrow \infty$,  $H_{th}$ tends to zero:
 \be
\lim _{\eta/s \rightarrow \infty} \frac{H_{th}}{H_i} = \frac{3\sqrt{10}}{(\eta/s)^2} \frac{m_{pl}}{\hat{w}^2 \sqrt{g_*}\pi H_i}
 \ee 
 This is a simple consequence of the redshifting of $H$ over the course of thermalization.
 
This in turn determines the thermalization time $t_{th}$ via \eqref{Hretre}, and the thermalization temperature $T_{th}$ via \eqref{HreTre}. The latter is given by, 
\begin{eqnarray}
\label{Tre}
&&T_{th} = \left(\frac{90}{\pi^2 g_*}\right)^{\frac{1}{4}} \times   \\ 
&&  \Bigg( H_i m_{pl}+ \frac{\sqrt{g_*} \pi}{6\sqrt{10}} \left(\frac{\hat{w} \eta}{s} \right)^2 H_i ^2 \left[ 1 - \sqrt{1 + \frac{12 \sqrt{10}m_{pl}}{(\hat{w} \eta/s)^2 \sqrt{g_*} \pi H_i}}\,\right] \Bigg)^{1/2}. \nonumber
\end{eqnarray}
We can relate this to number of e-folds of thermalization $N_{th}$, using ${\rm d}N = H {\rm d}t$, and assuming $w$ a constant, leads to
\be
N_{th} =\frac{1}{\hat{w}} \log \left[ 1 + \frac{H_i}{T_{th}}\hat{w}   \frac{\eta}{s} \right] ,
\ee
with $T_{th}$ given in terms of $\eta/s$ via equation \eqref{Tre}, and $\hat{w}$ in range $1\leq \hat{w} \leq 2$, as defined in equation \eqref{hatw}.

These results allow for $N_{th}$ and $T_{th}$ to be specified uniquely via $\eta/s$ and the energy scale of inflation $H_i$. This can be directly related to the scalar-to-tensor ratio \cite{Guzzetti:2016mkm}, 
\be
\frac{r}{0.01} \simeq \left( \frac{H}{2.7 \times 10^{13}\, {\rm GeV}}\right)^2 . 
\ee
Hence the thermalization process is itself sensitive to the tensor-to-scalar ratio.

\section{Thermalization time at Strong and Weak Coupling}

We are now in a position to make quantitative statements as to the thermalization time of the universe following inflation.  We note that the energy scale of the thermalization process is not a priori the inflationary energy scale, but instead is the energy of the decay products predominantly produced at (p)reheating. This is sensitive to the (p)reheating mechanism; whether it is broad or narrow resonance, or perturbative decays, and could be in IR modes or UV modes relative to $H$.

To compute the thermalization time, one needs the shear-viscosity-to-entropy ratio $\eta/s$. This is calculable via different methods, and at weak coupling we have \cite{Keegan:2015avk,Arnold:2003zc}
\be
\frac{\eta}{s}|_{\alpha_s \ll 1} = \frac{34.784}{\lambda^2 {\rm log} \left[ 4.789/\sqrt{\lambda}\right]} ,
\ee
where $\lambda$ is the `t Hooft coupling, $ \lambda = g^2 N_c $ for $N_c$ colors of gluon, which is related to the strong coupling constant $\alpha_s \equiv \frac{g^2}{4 \pi}$ by $\lambda = 4 \pi N_c \alpha_s$.  As a fiducial example, for $\lambda=0.1$ this evaluates to $\eta/s \simeq 10^3$. On the other hand, at strong coupling, $\eta/s$ is bounded:
\be
\frac{\eta}{s}|_{\alpha_s \gg 1} \gtrsim \frac{1}{4 \pi}.
\ee
as discussed below equation \eqref{etasbound}.

From this one can immediately deduce the AdS/CFT prediction for the thermalization time at reheating. It follows from \eqref{eq:tre}, as 
\be
N_{th} |_{\alpha_s \gg 1}= \frac{g_* ^{1/4}}{4 \sqrt{3 \pi} 10^{1/4}} \sqrt{\frac{H_i}{m_{pl}}} + \mathcal{O}\left(\frac{H_i}{m_{pl}}\right) .
\ee
This is much less than $1$ even for high scale inflation. Thus we see that thermalization at strong coupling is always \emph{cosmologically fast}, lasting for much less than one e-fold of expansion. 

Moreover, the thermalization temperature is, 
\be 
T_{th}|_{\alpha_s\gg1} \simeq  \left(\frac{90}{\pi^2 g_*}\right)^{\frac{1}{4}}  \sqrt{H_i m_{pl}} \left[ 1 + \mathcal{O}\left(\frac{H_i}{m_{pl}}\right)\right].
\ee
To leading order in $H_i/m_{pl}$ this is the standard result of instant reheating. It follows that reheating at strong coupling is in no danger from a low thermalization temperature.

On the other hand, to make predictions at weak coupling one must understand in detail the renormalization group flow of the strong coupling constant. In standard QCD, from the $\beta$-function for $SU(3)$ with $N_f=6$, the energy-scale dependence of $\alpha_s$ is given by,
\be
\alpha_s (\mu) = \frac{0.1187}{1 + 0.17\, \log \left[ \mu/M_Z\right]},
\ee
with $M_Z = 91.2 \, {\rm GeV}$. At very high energies, e.g. $\mu=m_{pl}$, this is only as small as $0.0198$.  However, the high-energy behavior of $\alpha_s$ is sensitive to new particles that emerge at high energies.  The inclusion of new gauge bosons accelerates asymptotic freedom, as the $\beta$-function scales with $N_c$.  Additionally, at energies near the Planck one expects quantum gravity effects to become important, which may radically alter the evolution of the effective gauge coupling.

\begin{figure}[h!]
\begin{center}
\includegraphics[width=0.49\textwidth]{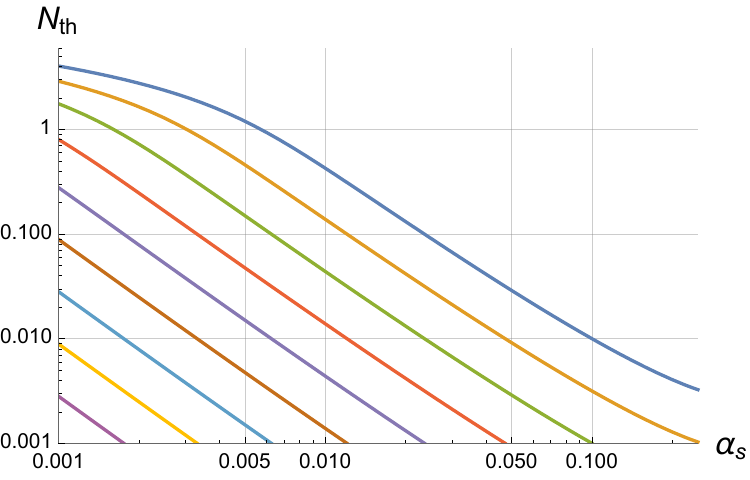}
\end{center}
\caption{Number of e-folds of thermalization, using the weak coupling prediction for $\eta/s$, as a function of the strong coupling constant, at varying energy scales of inflation, from top to bottom, $H_i =10^{13}, 10^{12},..., 10^{5} {\rm GeV}$, corresponding to tensor-to-scalar ratio $r=.014,...,10^{-19}$. We fix $g_*=100$ and $w=1/3$. }
\label{fig:Nth}
\end{figure}

With these possibilities in mind, here we consider the number of e-folds of thermalization at strong and weak coupling. See Figure \ref{fig:Nth}.  For simplicity here we identify the energy of scatterings $\mu$ with the energy scale of inflation, and we fix $w=1/3$.  From this plot one can appreciate that the thermalization time is sensitive to the energy scale of inflation, and thermalization generically takes longer at weak coupling and/or in large field inflation, taking of order or less than an e-fold of expansion. On the other hand, for both low-scale inflation and moderate coupling, thermalization takes much less than an e-fold.

\section{Observation}

The analysis thus far is an important consistency check on the inflationary universe scenario. It is also important to understand the quantitative impact of thermalization on the observables of inflation. 

The canonical observational probe of post-inflationary evolution, e.g.~reheating, is a shift in $n_s$ and $r$  \cite{Martin:2010kz,Munoz:2014eqa,Gong:2015qha,Cook:2015vqa,Dai:2014jja,Drewes:2015coa,Lozanov:2016hid,Easther:2013nga,Dai:2014jja,Martin:2016oyk,Drewes:2015coa,Lozanov:2016hid,Ji:2019gfy,Kabir:2016kdh}. This occurs solely due to the modified expansion history of the universe, and not the detailed microphysics. A period of expansion lasting $\Delta N$ e-folds and with equation of state $w_{\Delta}$ is completely degenerate with a shift in the number of e-folds before the end of inflation that the CMB pivot scale exited the horizon, by an amount \cite{Martin:2010kz,Cook:2015vqa},
\begin{equation}
\Delta \bar{N} = \frac{1}{4}(1- 3 w_{\Delta}) \Delta N .
\label{eq:deltabarN}
\end{equation}
Applied to reheating, $w_{\Delta}$ and $\Delta N$ correspond to the equation of state $w_{re}$ and number of e-folds of reheating $N_{re}$. The equation of state $w_{re}$ may deviate from $1/3$, and in full generality need only satisfy $w>-1/3$ (corresponding to the end of inflation). In practice, it can be treated as a free (bounded) parameter that is to be constrained by data.

Eq.~\eqref{eq:deltabarN} completely encodes the reheating dependence of inflationary predictions.  In particular, the Hubble constant at the moment a fluctuation of wavenumber $k$ exits the horizon $H_k$, from which one can derive $n_s$ and $r$,  is given by \cite{Cook:2015vqa},
\be
\log \left( \frac{H_k}{V_{end}^{1/4}}\right)=  N_k + \Delta \bar{N}- 61.6,
\ee
where $N_k$ is the number of e-folds before the pivot scale exits the horizon, and for simplicity the numerical factor of 61.6 accounts for the spin-degrees of freedom (fixed in the above to be 100) and the CMB pivot scale, taken to be $.05 \, {\rm Mpc}^{-1}$.

This determines $n_s$ and $r$ in a model-dependent and -independent fashion respectively. For polynomial potentials, $V=m^{4-p}\phi^p$, one finds \cite{Cook:2015vqa},
\be 
H_k = \pi M_{pl} \sqrt{\frac{4 \pi A_s}{p+2} (1-n_s)} \;\; ,\;\; r = \frac{2 H_k ^2}{\pi^2 M_{pl}^2 A_s}.
\ee  
As a numerical example, consider $m^2 \phi^2$ inflation, which is not a good fit to CMB data, but as a toy model is suitable for our purposes. The modification to the spectral index from a non-thermal phase is \cite{Cook:2015vqa},
\be
n_s = 1 - \frac{8}{230.6 - (1 - 3 \bar{w}_{\Delta})N_{\Delta} } .
\ee
The modification to the spectral index indeed vanishes if $w=1/3$, since in this case, the expansion history of the universe is unchanged from that in standard cosmology.

At energies far above the QCD scale, the quark gluon plasma is approximately conformal. The stress tensor is therefore traceless, $T_{\mu} ^\mu =0 $. This in turn implies that $\rho-3 p=0$ and hence $w=1/3$. Thus the thermalization process at sufficiently high energies will have no impact on the spectral index of primordial perturbations.

At lower energies, the situation is more subtle. We again emphasize that the energy scale of the thermalization process, $\mu$, is \emph{not} the inflationary energy scale, but instead the energy of the decay products predominantly produced at (p)reheating. This energy scale could be far below the inflationary scale, e.g. super-horizon modes produced via broad parametric resonance, or far above the energy scale, produced by perturbative decays of the inflaton.

The equation of state of the quark gluon plasma can be computed using lattice QCD, and at temperatures around $10^2$-$10^3$ MeV can differ substantially from $1/3$, see e.g.~\cite{Borsanyi:2016ksw}. For the present analysis it is sufficient to consider $0< w < 1/3$; the former bound arising from the limit of a collection of non-relativistic particles and the latter the high-temperature limit. In this case, using the previous result that $N_{th}\lesssim 1$, we find,
\be
\Delta \bar{ N} < \frac{1}{4} .
\ee
This shifts $n_s$ by at most the level of $1$ part in $10^4$; e.g. for $N_{th}=1$ and $w=0.1$, the shift in $n_s$ is $\delta n_s=3.9 \times 10^{-5}$. This is two orders of magnitude below the sensitivity of CMB S4 \cite{Hanany:2019lle} and Simons Observatory \cite{Ade:2018sbj}.

However, this is not the whole story: The non-Abelian gauge fields themselves will source gravitational waves. Gravitational waves are in general sourced by anisotropic stress, as encoded in the equation of motion,
\begin{equation}
h_{ij} '' + 2 \mathcal{H} h_{ij} ' - \nabla^2 h_{ij} = \frac{2}{M_{ Pl} ^2} T_{i j} ^{TT},
\end{equation}
 where $T_{ij} ^{TT}$ is the spatial off-diagonal components of the stress-tensor, or more formally, the transverse traceless projection of the spatial stress tensor.   The spatial stress tensor is given in full generality as,
\be
\label{Tij}
 T^{ij}_{ }  =
 \bigl( p - \zeta \nabla\cdot \vec{v} \bigr) \, \delta_{ }^{ij} \nonumber- \eta \bigl( \partial^i v^j_{ } + \partial^j v^i_{ }
 - \frac{2}{3} \delta^{ij}_{ } \nabla\cdot \vec{v} \bigr)
 \; .
\ee
One can easily see that it is $\eta$ which will play a key role in generating gravitational waves.

 The sourcing of gravitational waves from a non-Abelian plasma was studied in detail in \cite{Ghiglieri:2015nfa}. The sourcing of  gravitational wave spectra is,
\bea
 \frac{{\rm d}\rho^{ }_{\rm GW}}{{\rm d}t\, {\rm d}^3\vec{k}} 
 \; = \; &&
 \frac{4\pi G}{(2\pi)^3} 
 \sum_{\lambda} 
 \epsilon_{ij,\vec{k}}^{{\rm TT}(\lambda)}
 \epsilon_{mn,\vec{k}}^{{\rm TT}(\lambda)*}   \\
 \nonumber && \int_{\mathcal{X}}
  e^{i\mathcal{K}\cdot\mathcal{X}} 
 \big\langle \,
    T_{ }^{ij} (0) \,
    T_{ }^{mn} (\mathcal{X}) \, 
 \big\rangle^{ }_{ } \; ,
 \eea
where the $\epsilon_{ij}^{{\rm TT}}$ are polarization tensors. Evaluating the sum over polarizations, and choosing 
$\vec{k} = k\, \vec{e}_3$,  leads to
\be
 \frac{{\rm d}\rho^{ }_{\rm GW}}{{\rm d}t\, {\rm d}\ln k} 
 \; = \; 
 \frac{8 k^3}{\pi m_{\rm Pl}^2 }  \int_{\mathcal{X}} e^{ik(t-z)}
 \big\langle \, T^{ }_{12} (0) \,T^{ }_{12} (\mathcal{X}) \, 
 \big\rangle^{ }_{ } \;. 
\ee 
Utilizing the stress tensor, one can simplify this on large scales,
\be
 \lim_{k \to 0}
 \int_{\mathcal{X}} 
  e^{ik(t-z)}
 \Bigl\langle
  \frac{1}{2} 
 \bigl\{ T^{ }_{12}(\mathcal{X}) , T^{ }_{12}(0) 
 \bigr\}
 \Big\rangle^{ }_{ }
 = 2\, \eta\, T ,
 \;
\ee
which gives the power in gravitational waves on large scales as,
\be
\frac{{\rm d}\rho^{ }_{\rm GW}}{{\rm d}t\, {\rm d}\ln k} 
 \; = \; 
 \frac{8 k^3}{\pi m_{\rm Pl}^2 } 2\, \eta\, T.
\ee
We can relate this to $\eta/s$ using $s=(2\pi^2/45) g_{*s} T^3$ in a radiation dominated universe, with $g_{*s} \simeq g_*$, 
as well as $H^2 =(\pi^2/30) g_* T^4/3{m_{pl}}^2$, to find,
\be
\frac{{\rm d}\rho^{ }_{\rm GW}}{{\rm d}t\, {\rm d}\ln k}  = \frac{64 k^3}{\pi}  \left(\frac{\eta}{s}\right)  H^2 .
\ee
The net production of gravitational waves at thermalization is then given by an integral over the thermalization time period. 

As an estimate, guided by the results of the previous section, we consider the limit of fast thermalization, $N_{th} \ll 1$. We change the time coordinate to the number of e-folds,
\be
\frac{{\rm d}\rho^{ }_{\rm GW}}{{\rm d}N\, {\rm d}\ln k}  = \frac{64 k^3}{\pi} \left(\frac{\eta}{s}\right)  H,
\ee
and approximate thermalization as an instantaneous process. The produced gravitational waves are,
\be
\frac{{\rm d}\rho^{ }_{\rm GW}}{{\rm d}\ln k}  = \frac{64 k^3}{\pi}  \left(\frac{\eta}{s}\right)  H_{th}.
\ee
The signal observed \emph{today} is related to the above by a simple redshifting by $a(t)^{-4}$:
\be
\frac{{\rm d}\rho^{ }_{\rm GW}}{{\rm d}\ln k} (t_0) = \left(\frac{a(t_{th})}{a(t_0)}\right)^4 \frac{{\rm d}\rho^{ }_{\rm GW}}{{\rm d}\ln k} (t_{th}).
\ee
The redshift factor can itself be re-expressed in terms of the thermalization temperature, by relating to known quantities at matter-radiation equality.

The resulting present day spectrum of gravitational waves is given by,
\be
\frac{{\rm d}\rho^{ }_{\rm GW}}{{\rm d}\ln k} (t_0) = 2.2 \times 10^{-12} k^3  g_* \left(\frac{\eta}{s}\right)  \frac{{\rm eV}^4}{H_{th} m_{pl}^2}.
\ee
From this one can appreciate that the redshifting of gravitational waves has flipped the dependence on $H_{th}$: high-scale thermalization will lead to a lower amount of gravitational waves in the present universe.

The gravitational wave spectrum is heavily blue tilted, and thus suppressed on large scales relative to a scale-invariant spectrum. This is a feature common to all post-inflation gravitational wave production mechanisms (see \cite{Guzzetti:2016mkm} for a review), wherein the signal is peaked at a characteristic length scale and decays on scales larger than this. This makes the effect unimportant for CMB B-modes, but opens the possibility for observing it with interferometer experiments. The quantity of interest is, see e.g. \cite{Guzzetti:2016mkm},
\be
\Omega_{\rm GW}(f) = \frac{1}{\rho_c} \frac{{\rm d} \, \rho_{\rm GW}}{ {\rm d \, ln} f},
\ee
where $\rho_c$ is the critical energy density of the universe today.  We translate from $k$-mode to frequency space via $f=k/2\pi$,  following the unit conversion  $k/{\rm Mpc}^{-1} = 6.5 \times 10^{14} \, f/{\rm Hz}$. From this we find,
\be
\Omega_{\rm GW}(f) = 1.6 \times 10^{-108} g_* \left(\frac{\eta}{s}\right)  \left(\frac{{\rm GeV}}{H_{th}}\right)  \left(\frac{f}{\rm Hz}\right)^3 .
\ee
The extremely small prefactor prevents this signal from reaching any appreciable level. This is simply a reflection of the hierarchy of scales in the problem, namely the Hubble scale at thermalization and the scales of interest to LIGO.

The current LIGO constraint \cite{TheLIGOScientific:2016dpb} on a stochastic gravitational wave background is given by,
\be
\Omega_{\rm GW}(f) < 1.7 \times 10^{-7} \;\; , \;\; f\sim 20 - 86 \,{\rm Hz}.
\ee 
Even in the most optimistic case, saturating the BBN bound $H_{th}\sim {\rm MeV}$, and assuming $(\eta/s)\sim 10^3$, $g_*\sim 10^2$, the gravitational wave signal from thermalization is well below the observational bound.\\

\section{Discussion}

In this work we have applied and generalized results from the thermalization of non-Abelian gauge theories, originally developed to understand the formation of quark gluon plasma at heavy ion colliders, to the epoch of reheating after cosmic inflation. We derived a relation between the number of e-folds of the thermalization process, the shear-viscosity-to-entropy ratio $\eta/s$, and the scalar-to-tensor ratio $r$. We find that thermalization generically completes within less than a single e-fold of expansion, without significantly modifying the inflationary predictions for $n_s$ and $r$. This indicates that the predictions of inflationary models are robust to the physics of thermalization.

It is imperative to study further possibilities for observing the microphysics of the formation of the primordial quark gluon plasma. For example, it is known that thermalization of the quark gluon plasma can be delayed by a long-lived phase nearby a non-thermal fixed point \cite{Berges:2008wm}; it would be interesting to consider the cosmological realization of this, and the impact on $n_s$ and $r$. It is also possible that there could be a phase transition during the thermalization process itself, which would further source gravitational waves. We leave these interesting possibilities for future work.

Finally, we have not endeavored to consider the detailed interplay of thermalization and preheating. Holographic methods have been applied to preheating in \cite{Cai:2016sdu,Cai:2016lqa}. It would be interesting to consider the gravitational wave signal once these effects are included.

\acknowledgements
EM thanks an anonymous referee for insightful comments and helpful suggestions, as well as Stephon Alexander, Robert Brandenberger,  Elisa Ferreira, Alan Guth, Colin Hill, David Kaiser, Aleksi Kurkela, Jacquelyn Noronha-Hostler, Jorge  Noronha, Brandon Melcher, Paul Romatschke, Vincent Vennin, Scott Watson, and William Allen Zajc, for helpful comments.

\newpage

\bibliographystyle{JHEP}
\bibliography{QGP-refs}

\end{document}